\documentclass[12pt]{article}
\usepackage{a4,amssymb}
\usepackage{graphicx}

\newcommand{\xbj}{x_{\mbox {\tiny Bj}}}

\begin{document}
\begin{center}
{\Large {\bf Real and Virtual Compton Scattering in a Regge Approach}}
\end{center}

\vspace{1cm}

\begin{center}
F. Cano and  J.-M. Laget
\end{center}

\begin{center}
CEA-Saclay, DAPNIA/SPhN \\
F-91191 Gif Sur Yvette Cedex, France
\end{center}
 
\vspace{2cm}

\begin{abstract}
{\small We study Real and Deeply Virtual Compton Scattering in a model
based on Regge trajectories and two-gluon exchange. In
the kinematic regime of current experiments, the hadronic component of
the outgoing real photon plays a major role. We analyze the spin structure of
Compton scattering at large momentum transfer and give predictions for
several spin asymmetries. In the DVCS channel, a fairly good agreement 
is obtained for the recently measured 
beam spin and charge asymmetries.
} 
\end{abstract}

\vspace{2cm}
\noindent 

\vspace{1cm}
\noindent 
fcano@cea.fr \\
jlaget@cea.fr (corresponding author)

\vfill
\noindent
{\it Preprint} {\bf CEA-SPhN-}
 
\newpage

{\bf{1. Introduction.}} Exclusive real and
virtual Compton scattering are the cleanest channels for the investigation 
of the structure of hadronic matter. They involve two electromagnetic 
couplings and leads to more complete information than other processes like 
deep inelastic scattering or elastic form factors.

It was recently established that, for high virtuality $Q^2$, the amplitude of
 deeply virtual Compton scattering (DVCS) can be factorized in a hard part,
which can be calculated in pQCD, and a soft universal non-perturbative
part which is parameterized in terms of Generalized Parton
Distributions (GPDs) (for a recent review see \cite{GOEKE01} and
references therein). The existence of a hard scale ($Q^2$) leads to
the dominance of the handbag-type diagrams, where the two photons have a
pointlike coupling with the quarks of the target. 
Since DVCS is a process of order
$\alpha_{\mbox {\tiny em}}^3$, cross sections are very small and its
experimental determination is a very difficult task. So far, absolute
cross sections have been released only by HERA in the small $\xbj$ region
\cite{H101}. The beam spin
asymmetry is more accessible and has already been measured at JLab
\cite{CLAS01} and HERMES \cite{HERMES01}. The size of this asymmetry
is an indirect determination of the GPDs (more precisely, the
imaginary part of the VCS amplitude). Theoretical
expectations based on current modelisations of GPDs are compatible with
both HERMES and JLab data but the general trend is an overestimate of
the experimental points. 

However, in such a range of low momentum transfert  the coupling of the point
like component of the final real photon might not be dominant: the contribution 
due to its hadronic component is not negligible. Indeed, the coherence
length, or lifetime of the hadronic component, of a real photon of, let say,
4 GeV is of the order of 3 fm, larger than the size of the nucleon. This
conjecture  has been verified by  measurements at Cornell
\cite{CORNELL81} and more 
recently by HERMES~\cite{HERMES99}. The emission of the real photon occurs 
through the formation of a vector meson intermediate state. The photon 
production amplitude can simply be obtained from the vector meson production 
amplitude\footnote{The photoproduction 
of an isovector
state dominates over the isoscalar channel, so that we can keep only
the $\rho$ channel and neglect the $\omega$ channel} 
just by multiplying it by the conversion factor 
$\sqrt{4 \pi\alpha_{\mbox {\tiny em}}}/f_V$, where $f_V$ is the radiative 
decay constant of the vector meson~\cite{CANO02} (see also
\cite{DONNACHIE01}). We already showed~\cite{CANO02} that
this is supported by available data  on
cross sections of 
$\rho$-photoproduction~\cite{CLAS01RHO} and wide angle Compton
scattering (WACS) at $E_\gamma = 4$ GeV \cite{CORNELL79}.

\begin{figure}[t]
\begin{center}
\includegraphics[scale=0.6]{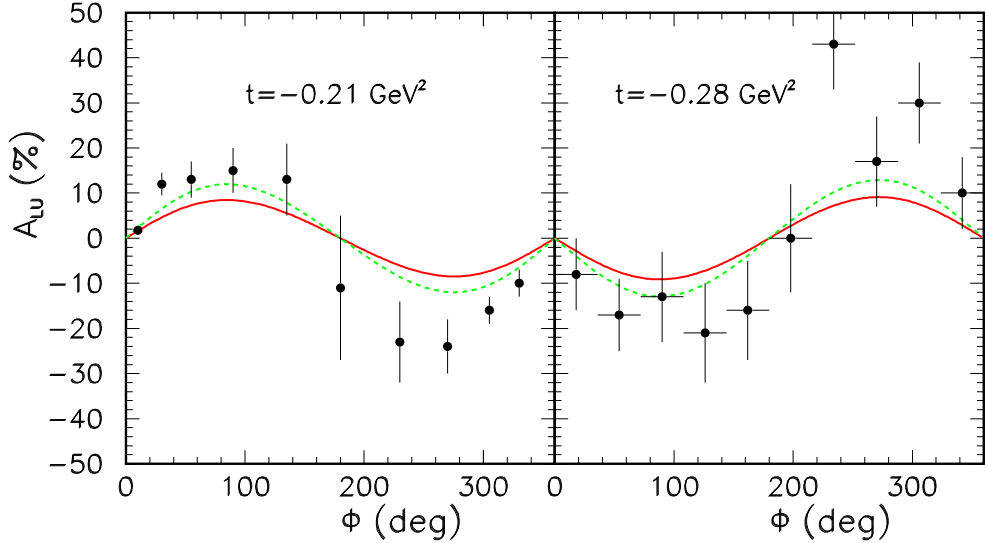} 
\end{center}
\caption{Azimuthal dependence of the beam-spin asymmetries in DVCS. 
Data taken from CLAS
\protect\cite{CLAS01} (left, $\xbj=0.19$, $Q^2=1.25$ GeV$^2$,
$E_{e^-}=4.25$ GeV) and HERMES \protect\cite{HERMES01}
(right, $\xbj=0.11$, $Q^2=2.6$ GeV$^2$,
$E_{e^+}=27.6$ GeV).The full and dashed curves correspond respectively to the
canonical and the renormalised conversion factor.}
\label{beam_as}
\end{figure}

In this letter, we extend to spin observables the predictions of 
this effective model
based on the interaction of the hadronic component of the photon with
the proton through the exchange of mesons (Regge trajectories) and of two  
non-perturbative gluons. It turns out, 
Fig.~\ref{beam_as}, that it reproduces fairly well the beam
spin asymmetries $A_{LU}$ observed in DVCS, not only  their $\sin \phi$ 
dependence, which
is just a consequence of the dominance of the helicity conserving  amplitude for
absorbing a transverse virtual photon \cite{DIEHL97}, 
but also their magnitude. 
In the kinematical range accessible by the
present generation of experimental facilities, we have not yet reached the
asymptotic regime where the point like coupling of the photon dominates 
\footnote{In the previous version of this report (Phys. Lett. {\bf B551} (2003)
317) the Compton amplitude was overestimated by a factor 1.95, due to a numerical
 mistake in the evaluation of the conversion factor. We thank M. Diehl for
pointing out this inconsistency. }.

{\bf{2. The theoretical framework.}}
Let us briefly recall the main ingredients of the model for
vector meson production. At high
energies the interaction of the photon with the proton occurs through
the exchange of Regge trajectories, which represent an economical way
to take into account meson (or $q\bar{q}$) exchange. In addition,
gluon exchange is also allowed and adds up to the Regge (or quark)
exchange. We refer to~\cite{LAGET00} for the expression of the various gluon,
meson and baryon exchange amplitudes. We give only the expression of the
$\sigma$ meson exchange gauge invariant amplitude, which was not given there. 
For $\rho$ production, it  is much larger that $\pi$ exchange, 
and its vector part takes the form~\cite{SOYEUR96}:

\begin{equation}
\vec{J} \cdot \vec{\epsilon_\gamma} = i e \frac{g_{\rho \sigma \gamma
}}{m_\rho} g_{NN \sigma} \bar{u}(\vec{p}\, ',s') u(\vec{p} ,s)
{\cal{P}}_\sigma \left[ q (p-p') \vec{\epsilon_\rho}^{\, *} \cdot
\vec{\epsilon_\gamma} - q \epsilon_\rho^{\, *}
(\vec{p} -\vec{p'})\cdot \vec{\epsilon}_\gamma\right]  ,
\end{equation}

\noindent where $q$ refers to the momentum of the incoming photon, $p$
and $p '$ to the momenta of the initial and final proton. Contrary to
Ref.~\cite{SOYEUR96} we use the Regge propagator ${\cal{P}}_\sigma$  
whose expression is~\cite{GUIDAL97}:

\begin{equation}
{\cal{P}}_\sigma = \left( \frac{s}{s_0} \right)^{\alpha_\sigma(t)} \frac{\pi
\alpha_\sigma '}{\Gamma(1+\alpha_\sigma(t))} \frac{e^{-i \pi
\alpha_\sigma(t)}}{\sin (\pi \alpha_\sigma(t))} .
\end{equation} 

The reference scale $s_0$ is chosen to be 1 GeV$^2$ and the Regge
trajectory is $\alpha_\sigma(t) = \alpha_\sigma^0 + \alpha_\sigma ' t$
with a slope $  \alpha_\sigma ' = 0.7$ GeV$^{-2}$ and an intercept
given by $ \alpha_\sigma^0 =  -\alpha_\sigma ' m_\sigma^2 = -0.175$. 
The coupling
constants are $g_{\rho \sigma \gamma}= 1$ and $g^2_{NN \sigma}/4
\pi = 15$. They fall in the range of values which can be deduced from the
analysis of the radiative decay width $\rho\rightarrow \gamma
(\pi\pi)_S$~\cite{RPP00} and of nucleon-nucleon scattering~\cite{MACHLEIDT89}.
Since the $\sigma$ meson is a representation of the two pion $S$-wave continuum,
their is an inherent uncertainty in their definition, and their product has been
determined by fitting~\cite{LAGET00} the low energy $\rho$ photoproduction data.
As usual, an hadronic form factor,
$F_{\sigma}(t)=(\frac{\Lambda^2-m_\sigma^2}{\Lambda^2-t})^4$ with
$\Lambda=2$ GeV, is used at
the $NN \sigma$ vertex. In addition, 
at low momentum transfer and moderate energies, also the $f_2$ meson exchange 
is important. 

	At large momentum transfer Regge trajectories saturate
and become $t$ independent. Saturating trajectories \cite{GUIDAL97} 
represent an
effective way to take into account the formation of a meson through
the exchange of a hard gluon, and the model agrees remarkably well with
data~\cite{CLAS01RHO}, in the region around 90 degrees. 
Finally, as the kinematical
boundary in $t$ is reached, the u-channel exchange of $N$ and $\Delta$
becomes the dominant mechanism. A more quantitative analysis of 
the relative weight of each 
contribution in different kinematical regimes can be found in
\cite{LAGET00}.

{\bf{ 3. Real Compton Scattering.}} 
In Ref. \cite{CANO02} we have shown how the various
 meson photoproduction channels at large momentum transfer single out
and calibrate the various
ingredients of this this model. When scaled by
the factor $\sqrt{4 \pi \alpha_{\mbox {\tiny em}}}/f_V$ the 
$\rho$-photoproduction amplitudes\footnote{We have retained only the
transverse $\rho$ polarization.} lead to a "parameter free" prediction of the
Compton scattering amplitudes, giving access not only to the cross
sections~\cite{La02} but also to the various spin observables.

For instance, the energy dependence at fixed angle for
$\gamma p \rightarrow p \rho$ at 90$^\circ$ is compatible with a
$s^{-7}$ behaviour. The $\gamma p \rightarrow p \gamma$ cross section
also follows this behaviour because the $\rho \gamma$ coupling
does not introduce any extra power in $s$. Cornell data \cite{CORNELL79}
at fixed angle were fitted to a power $s^n$, with $n=7.1\pm 0.4$ at
90$^\circ$. Models based on soft overlap \cite{RADYUSHKIN98} also
predict an approximate $s^{-7}$ behavior at these angles. This is at
variance with the pQCD counting rules which lead to a  $s^{-6}$ 
power-law. Nonetheless, more precise data is needed in
order to settle with greater accuracy the energy dependence.

Polarization observables impose futher constraints. Of
particular interest are the asymmetries which are being measured at
JLab. The longitudinal polarization transfer is defined as

\begin{equation}
A_{LL} \frac{d\sigma}{dt} = \frac{d\sigma(\uparrow \uparrow)}{dt} -
\frac{d\sigma(\uparrow \downarrow)}{dt} ,
\end{equation} 

\noindent where the first arrow indicates the (positive) polarization
of the incident photon and the second one refers to the helicity of the
recoiling proton.

In analogous way, we can define the transverse polarization transfer
of the outgoing proton $A_{LT}$ as\footnote{Following
\cite{BOURRELY80,HUANG02}, the direction of the normal (to the
scattering plane) polarization is defined as $\vec{N} = \hat{q} \times
\hat{p} '$ and the transverse polarization as $\vec{T} = \vec{N} \times
\hat{p} '$, the $\hat{z}$-direction taken along the incoming photon.}:

\begin{equation}
A_{LT} \frac{d\sigma}{dt} = \frac{d\sigma(\uparrow \rightarrow)}{dt} -
\frac{d\sigma(\uparrow \leftarrow)}{dt} ,
\end{equation}

\noindent and the induced polarization in the normal plane
$P_N$ for unpolarized photons.

\begin{figure}
\begin{center}
\includegraphics[scale=0.5]{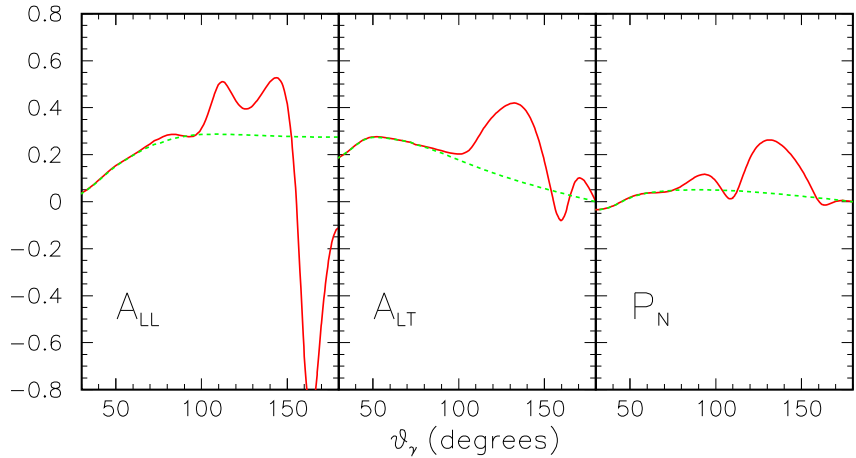} 
\end{center}
\caption{Longitudinal (left), transverse polarization transfer
(center) and induced polarization (right) in Compton
Scattering at $E_\gamma = 4$ GeV. 
Dashed lines are the contribution of Regge
Exchange in the $t$-channel. Solid lines are the
final results, which include $u$-channel exchanges.}
\label{real_compton}
\end{figure}

In Figure~\ref{real_compton} we present our results for these three polarization
transfers. Our predictions are quite close the  ones provided by the
soft overlap mechanism~\cite{HUANG02,DIEHL99PL} and have opposite sign 
with respect to the pQCD one~\cite{VANDERHAEGHEN97}. Preliminary data 
from JLab~\cite{NATHAN02}  ($A_{LL} \approx 70
\%$ at 120$^{\circ}$) confirm our conjecture and rule out the pure asymptotic
hard scattering approach. Our curve lies slightly below this experimental point.

For $A_{LT}$ our predictions are also similar to the ones of the soft
overlap approach, both in sign and in magnitude. Our prediction for the 
induced normal polarization is rather small,
$P_N \simeq 20$ \% at most, as in the handbag approach where it is a NLO
effect (order $\alpha_s$). 

	We can trace back the origin of these asymmetries by writing
the polarization transfer in a given direction $\hat{n}$ as~\cite{la94}:

\begin{equation}
A \frac{d\sigma}{dt} = 2 Im\left[ W_{xy}(\hat{n}) - W_{yx}(\hat{n})
\right] , 
\end{equation}

\noindent where 

\begin{equation}
W_{\mu \nu} (\hat{n}) = \sum_{m_1,m_2,m_2'} \langle m_1 | J_\nu^\dag |
m_2 ' \rangle \langle m_2 | J_\mu | m_1 \rangle (\vec{\sigma} 
\cdot \hat{n})_{m_2',m_2}
\end{equation}

\noindent and $J_{\mu}$ is the current which couples to the initial
photon. To have a non-vanishing polarization transfer we first need
phases in the amplitudes which are provided by the Regge
propagators. Second, a helicity flip in the proton sector is
required. Neither the $\sigma$, $f_2$ nor two-gluon exchange amplitudes
can flip the helicity of the proton. Only the $\pi$-exchange is able
to provide this flip. The dominant contribution to $A_{LL}$ for angles
$\leq $ 120$^\circ$ comes from the interference of the
$\sigma$-exchange with the $\pi$-exchange. The inferference of the
other helicity conserving amplitudes ($f_2$, two-gluons) with the
$\pi$ represents only a small correction. The contribution to $P_N$
comes from the $\pi-\sigma$ interference, but it is proportional to
the factor ($\frac{\vec{p} \times \vec{p} '}{(p^0 + m_N)(p'^0+m_N)}$)
which is small compared to the factors which takes part in the other
asymmetries. Finally, for angles larger than $\approx$ 120$^\circ$ the
asymmetries are completely dominated by baryon exchanges in the $u$
channel.

{\bf{Deeply Virtual Compton Scattering.}} In a similar way, the DVCS 
amplitude can be deduced from the electroproduction amplitude of a 
(transverse) vector meson.
	 
\begin{figure}
\begin{center}
\includegraphics[scale=0.6]{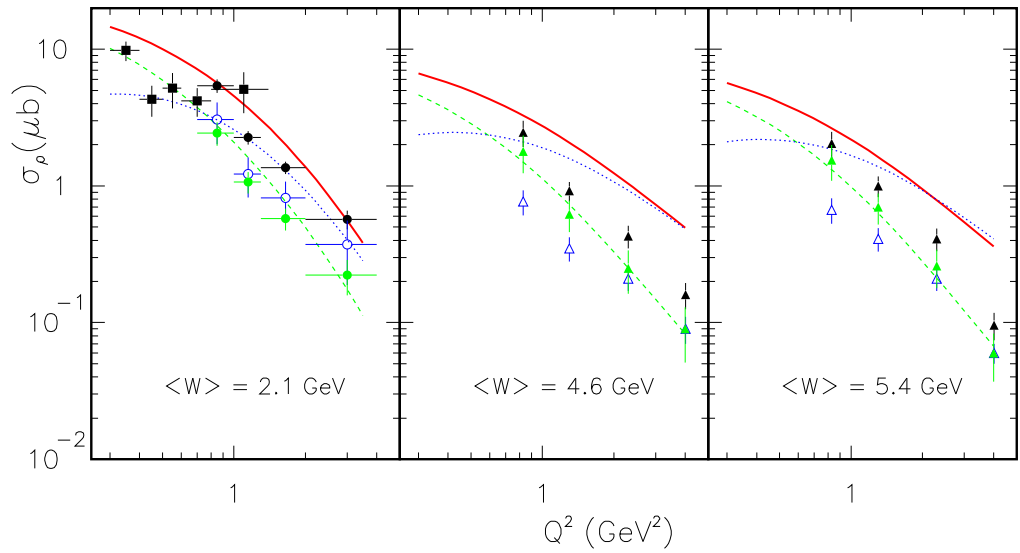} 
\end{center}
\caption{$Q^2$ dependence of the longitudinal (dotted lines) and
transverse (dashed lines) $\rho$ production cross section for several
values of W. Full lines: total cross section taking $\langle \epsilon
\rangle=0.95, 0.85, 0.72$ for $W=2.1, 4.6, 5.4$ GeV
respectively. Experimental data from
\protect\cite{DESY76} (boxes) and \protect\cite{CORNELL81} (circles) 
at $W=2.1$ GeV and from \protect\cite{HERMES00} at the other energies.
Black filled symbols: total, open symbols: longitudinal, faded
symbols: transverse cross section.}
\label{electrons}
\end{figure}

Let us first check the validity of the proposed model for meson 
electroproduction. The first difference with respect to the WACS case 
is that now we are interested in angular distributions at small angles 
or in the integrated  cross section which is essentially given by the 
low $-t$ region. Moreover, for virtual photons one has to introduce
electromagnetic 
form factors in the Regge-exchange amplitudes. The relevant amplitudes 
for HERMES energies and below are the $\sigma$, $f_2$ and two-gluon 
exchanges. The $Q^2$ dependence of the two-gluon and $f_2$ exchange
contributions is built in the
corresponding amplitudes \cite{LAGET00}. Concerning the
$\sigma$-exchange we have observed that 
a good description of $\sigma_{\gamma_T^* p \rightarrow p \rho}$ can be 
achieved with a monopole form factor $(1+Q^2/\Lambda^2)^{-1}$ with 
$\Lambda^2=0.46$ GeV$^2$ (this is in line with a VDM description of the
$\gamma\rho\sigma$ coupling). In Figure~\ref{electrons} we show the transverse, 
longitudinal 
and total cross sections for $\rho$-electroproduction at three different 
energies relevant for JLab and HERMES. We see that $\sigma_T$ is very 
well reproduced in these cases, though $\sigma_L$ is clearly overestimated. 
This is a longstanding problem in models of vector meson electroproduction 
and proposed solutions are based on a picture where the production of the 
meson takes place through open $q\bar{q}$ pairs \cite{DONNACHIE01,MARTIN97}. 
Since our final goal is to apply the model to DVCS  we do not address that 
problem here. The mean value $\langle \epsilon\rangle$ of the virtual photon
polarization, corresponding to each experimental setting, has been used  to
compute the total cross section $\sigma_{tot}= \sigma_T + \langle
\epsilon\rangle \sigma_L$.

\begin{figure}
\begin{center}
\includegraphics[scale=0.6]{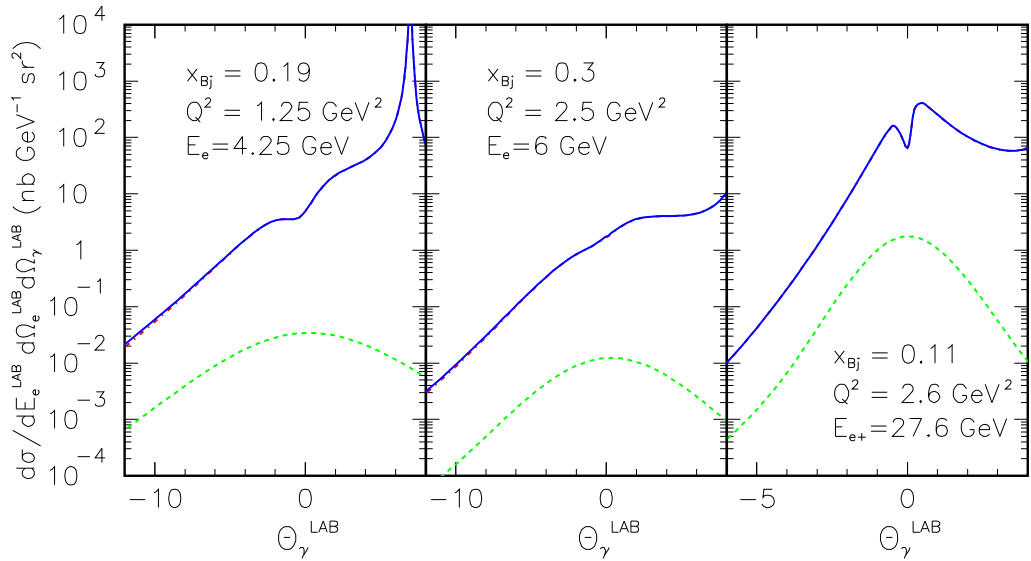} 
\end{center}
\caption{Differential cross section for Virtual Compton Scattering for
the kinematic relevant for JLab (left and center) and HERMES (right). 
Dashed-lines are the contribution of VCS and solid lines includes the 
Bethe-Heitler contribution.}
\label{X_sections}
\end{figure}

	The DVCS cross section is sensitive to the transverse amplitudes 
since the outgoing real photon has only transverse polarizations. 
The angular distributions for typical kinematics at JLab and HERMES 
are shown in Figure~\ref{X_sections}. The VCS cross section is roughtly ten times
smaller than predictions based on current modelisations of 
GPDs~\cite{VANDERHAEGHEN99}, and even 
for the region $\Phi=\pi$ where the Bethe-Heitler reaches its minimun, 
it overwhelms the VCS contribution. 

\begin{figure}
\begin{center}
\includegraphics[scale=0.5]{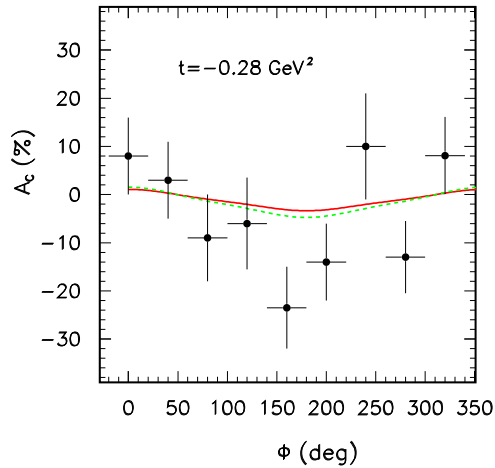}
\end{center}
\caption{Beam-charge asymmetry in DVCS for HERMES (preliminary data for 
\protect\cite{ELLINGHAUS02}) with the same kinematics as in Fig. 1. 
The full and dashed curves correspond respectively to the
canonical and the renormaized conversion factor.}
\label{charge_as}
\end{figure}

\begin{figure}
\begin{center}
\includegraphics[scale=0.5]{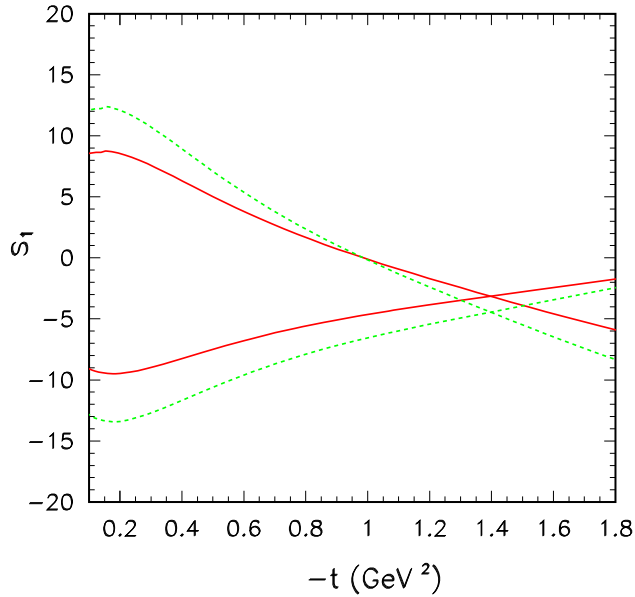}
\end{center}
\caption{Dependence of the leading term ($\sin \Phi$) with the momentum
transfer for CLAS (upper part) and HERMES (lower part). Kinematics as
in Fig.~\ref{X_sections}  (left and right panel
respectively). The full and dashed curves correspond respectively to the
canonical and the renormaized conversion factor.}
\label{s1}
\end{figure}

	Predictions for the measured beam asymmetries are also accordingly 
smaller than those based on GPDs, but in better agreement with
experiments (Fig. 1). The full curves correspond to the value  
$\sqrt{4 \pi\alpha_{\mbox {\tiny em}}}/f_V= 1./16.67$, as 
deduced~\cite{DONNACHIE01} from the radiative decay width of the $\rho$ meson.
However, this value underestimates, by a factor two, the real Compton scattering
cross  section at forward angle~\cite{leith78}, in the same range of momentum
transfert ($-t< 0.4$~GeV$^2$) and energy ($3<E_{\gamma}<16$~GeV) as covered in our
study. One possible reason is that the conversion factor may depend on the
virtuality of the photon: its canonical value is determined at the $\rho$ meson
pole, while we need its value in the space like region. If this conjecture is
correct, one may renormalize the conversion factor by $\sqrt{2}$ and get the
dashed line in Figs.~\ref{beam_as}, \ref{charge_as} and \ref{s1}. The difference between full
and dashed line gives a measure of the theoretical uncertainty.
This observable is sensitive to the 
interference between VCS and Bethe-Heitler (BH) diagrams and 
is proportional to the imaginary 
part of the VCS amplitudes. The $\sin \Phi$ dependence of the asymmetry is 
a general feature of helicity conserving interactions for spin-1 particles. 
A Fourier decomposition of our results for $A_{LU}$ gives $0.11 \sin(\Phi)+
0.006 \sin(2 \Phi)$ 
for CLAS and $-0.13 \sin(\Phi)- 0.020 \sin(2 \Phi)$ for HERMES kinematics. 

	The agreement with the experimental data is remarkable also for the 
beam charge asymmetry (Figure~\ref{charge_as}), that measures the real part 
of the amplitudes 
mentioned above. It has been argued in \cite{KIVEL01} that this observable is 
very sensitive to the D-term in the GPDs formalism. The D-term takes into 
account the scalar-isoscalar $q\bar{q}$ correlations in the proton. 
In our description, the $\sigma$-exchange seems to provide a good description 
of these correlations. 

In the HERA energy range, the two gluon exchange mechanism dominates (over the
Reggeon exchange mechanisms) and accounts for a sizeable part of the H1 DVCS 
cross section~\cite{H101} (0.8 nb, as compared to 4 $\pm$ 2
nb, at Q$^2$ = 5 GeV$^2$), confirming the findings of \cite{DONNACHIE01}.

{\bf{5. Discussion.}} The good results obtained for the observables 
measured so far for DVCS, starting from the $\gamma_T^* p \rightarrow p \rho$ 
amplitudes, supports the conjecture that the hadronic component 
of the photon dominates at currently available energies. There is however a crucial difference 
between VCS amplitudes evaluated from GPDs, which show a $1/Q$ behavior 
(up to logarithmic corrections), and the ones used in our model, which have 
a steeper $Q^2$ dependence due to electromagnetic form factors. 
This means that at larger 
$Q^2$ the hadronic mechanism will fade out and that the pointlike coupling 
of the photon is expected to take over. The exact
place where this happens is still under debate. Only experiments will answer
this question. For example, in a 
kinematics which will be reached at JLab when the beam energy is upgraded
up to 12 GeV ($Q^2=8$ GeV$^2$,  $\xbj=0.55$) our model predicts an almost 
vanishing beam spin asymmetry (less than  2\%), whereas models based on GPDs 
give results of the order of  30\% at $t=-1$  GeV$^2$ \cite{MOSSE01}.

Another feature that could help to reveal the dominance of the hadronic 
component of the photon is the $t$-dependence of the weight $s_1$ of the 
leading 
term ($s_1 \sin \Phi$) in $A_{LU}$. Due to the phase of the Regge propagator, 
the sign 
of the imaginary part of the amplitudes changes and consequently the 
sign of  $A_{LU}$ (Figure~\ref{s1}). This feature is sensitive to the
energy $s$ that controls the relative importance of different trajectories.
An analysis of the $t$-dependence of $A_{LU}$ at JLab 
energies would shed more light on the reaction mechanism.

In summary, the hadronic component of the outgoing photon dominates the cross
section and leads to a fair agreement with the  spin observables which have been 
determined so far for 
real Compton scattering  as well as deeply virtual Compton
scattering. It should be emphasized that the ingredients of the model
have been calibrated in meson photo and electroproduction channels ($\rho$,
$\omega$, $\phi$) and, therefore, predictions for WACS and DVCS
involve no additional parameter or refitting of the existing ones.  
More experimental data are needed, both in the vector meson production sector and in
the Compton scattering sector, to map out in a comprehensive way the behavior of
the hadronic component and find the best places to look for observables
associated with  the pointlike component of the photon in the initial as well
as the final states.  

\vspace{1cm}
We acknowledge F. Sabati\'e for help with Ref. \cite{MOSSE01}. 
This work is partially (FC) funded by European Commission IHP program 
(contract HPRN-CT-2000-00130).

\end{document}